\documentclass[12pt,draftclsnofoot,journal,onecolumn]{IEEEtran}
\usepackage{graphicx}
\usepackage{amsmath}

\usepackage{amsfonts}
\usepackage{amssymb}
\usepackage{array}
\newcommand{\PreserveBackslash}[1]{\let\temp=\\#1\let\\=\temp}
\newcolumntype{C}[1]{>{\PreserveBackslash\centering}p{#1}}
\newcolumntype{R}[1]{>{\PreserveBackslash\raggedleft}p{#1}}
\newcolumntype{L}[1]{>{\PreserveBackslash\raggedright}p{#1}}
\usepackage[usenames]{color}
\usepackage{colortbl,booktabs}
\usepackage{tabularx}
\usepackage{multicol}
\usepackage{booktabs}
\usepackage[Symbol]{upgreek}
\usepackage{subfigure}
\usepackage{bm}
\usepackage{cite}
\usepackage{balance}
\usepackage[ruled,vlined]{algorithm2e}

\usepackage{threeparttable}
\usepackage{amsmath}
\usepackage{dcolumn}
\usepackage{multirow}
\usepackage{amsfonts}
\newcolumntype{d}[1]{D{.}{.}{#1}}

\begin{document}

\bibliographystyle{IEEEtran} 
\title{Codebook Design and Beam Training for Extremely Large-Scale RIS: Far-Field or Near-Field?}

\author{Xiuhong Wei, Linglong Dai, Yajun Zhao, Guanghui Yu, and Xiangyang Duan
\thanks{X. Wei and L. Dai are with the Beijing National Research Center for Information Science and Technology (BNRist) as well as the Department of Electronic Engineering, Tsinghua University, Beijing 100084, China (e-mails: weixh19@mails.tsinghua.edu.cn, daill@tsinghua.edu.cn).}
\thanks{Y. Zhao, G. Yu, and X. Duan are with the Wireless Pre-Research Department, ZTE Corporation, Shenzhen 518038, China (e-mail: \{zhao.yajun1, yu.guanghui, duan.xiangyang\}@zte.com.cn).}
\thanks{This work was supported in part by the National Key Research and Development Program of China (Grant No. 2020YFB1807201) and in part by the National Natural Science Foundation of China (Grant No. 62031019).}
}

\maketitle
\vspace{-3em}
\begin{abstract}
     Reconfigurable intelligent surface (RIS) can improve the capacity of the wireless communication system by providing the extra link between the base station (BS) and the user. In order to resist the ``multiplicative fading" effect, RIS is more likely to develop into extremely large-scale RIS (XL-RIS) for future 6G communications. Beam training is an effective way to acquire channel state information (CSI) for the XL-RIS assisted system. Existing beam training schemes rely on the far-field codebook, which is designed based on the far-field channel model. However, due to the large aperture of XL-RIS, the user is more likely to be in the near-field region of XL-RIS. The far-field codebook mismatches the near-field channel model. Thus, the existing far-field beam training scheme will cause severe performance loss in the XL-RIS assisted near-field communications. To solve this problem, we propose the efficient near-field beam training schemes by designing the near-field codebook to match the near-field channel model. Specifically, we firstly design the near-field codebook by considering the near-field cascaded array steering vector of XL-RIS. Then, the optimal codeword for XL-RIS is obtained by the exhausted training procedure between the XL-RIS and the user. In order to reduce the beam training overhead, we further design a hierarchical near-field codebook and propose the corresponding hierarchical near-field beam training scheme, where different levels of sub-codebooks are searched in turn with reduced codebook size. Simulation results show the two proposed near-field beam training schemes both perform better than the existing far-field beam training scheme. Particulary, the hierarchical near-field beam training scheme can greatly reduce the beam training overhead with acceptable performance loss.

\end{abstract}


\section{Introduction}\label{S1}

Recently, reconfigurable intelligent surface (RIS) has been proposed to improve the capacity of the wireless communication system~\cite{ZhangEfficiency}. Specifically, RIS consists of a large number of reconfigurable elements (e.g., 256), which can be deployed between the base station (BS) and the user to establish an extra reflecting link. By properly reconfiguring the RIS elements, RIS can provide high reflecting beamforming gain with low cost and low consumption~\cite{RuiZhang19Beamforming}. The reliable RIS reflecting beamforming requires the accurate channel state information (CSI)~\cite{JunPrecoding,Zijian}. However, due to a large number of RIS elements, CSI acquisition is challenging for the RIS assisted system~\cite{CETutorial}.

There are two typical categories of methods for CSI acquisition, which are respectively the explicit CSI acquisition (i.e., channel estimation) and the implicit CSI acquisition (i.e., beam training). For the first category, the BS sends the pilot signals to the user through the RIS, and the user directly estimates the channel based on the received pilot signals~\cite{PartI}. Since the RIS element is usually passive, only the cascaded channel, i.e., the cascade of the channel from the BS to the RIS and the channel from the RIS to the user, can be estimated by least squares (LS)~\cite{Power'1} or minimum mean square error (MMSE) algorithm~\cite{Nadeem20DFT}. The high-dimensional cascaded channel estimation will lead to unaffordable pilot overhead in the RIS assisted system. In order to solve this problem, two types of low-overhead cascaded channel estimation schemes have been proposed~\cite{PartI}. On the one hand, some compressive sensing (CS) algorithms can be used to estimate the high-dimensional cascaded channel by leveraging the sparsity of the angular cascaded channel~\cite{JunCS,PartII}. On the other hand, the multi-user correlation is exploited to reduce the pilot overhead by considering that all users communicate with the BS via the same RIS~\cite{Wang20Correlation}. However, for this category of method, since RIS cannot preform the reliable reflecting beamforming before channel estimation, the received signal-to-noise ratio (SNR) is usually low. It is difficult for channel estimation to achieve the satisfactory channel estimation accuracy with the low received SNR.

The second category is beam training, where CSI can be obtained by estimating the physical directions of channel paths instead the entire channel. This beam training method has been widely considered in the existing 5G system, especially for millimeter-wave frequency~\cite{5GBT,5GCodebook}. Specifically, the BS and the user preform the training procedure through multiple directional beams (codewords) predefined in the codebook to search the optimal directional beam. After beam training, the physical directions of channel paths can be effectively obtained~\cite{XinyuBS}. Compared with channel estimation, beam training can directly achieve the reliable beamforming by the training procedure, which can avoid estimating the entire channel with the low received SNR. Recently, the beam training method has been extended to the RIS assisted system for CSI acquisition~\cite{RISBT,JunBT,DNBT}. The basic idea is that based on the cascaded array steering vector of the RIS cascaded channel, the codebook consisting of multiple RIS directional beams is firstly designed, and then the training procedure between the RIS and the user is performed to search the optimal RIS directional beam~\cite{RISBT}. By considering that the cascaded array steering vector is mainly determined by the angle differences at the RIS, the partial search based beam training scheme was further proposed to reduce the search complexity~\cite{JunBT}.


However, the existing codebook and beam training schemes may not be applicable any more with the increasing number of RIS elements. Specifically, the RIS assisted system is faced with the ``multiplicative fading" effect~\cite{VincentFading,PathLoss,RenzoRIS}, where the equivalent path loss of the BS-RIS-user reflecting link is the product of (instead of the sum of) the path losses of the BS-RIS link and RIS-user link. Thanks to low cost and low power consumption, more and more RIS elements are expected to deploy to compensate for the severe path loss~\cite{PathLoss}. RIS is more likely to develop into extremely large-scale RIS (XL-RIS) for future 6G communications, which will lead to the fundamental transformation of electromagnetic radiation field structure~\cite{Mingyao}. The electromagnetic radiation field can be divided into far-field region and near-field region~\cite{RayDistance}, which are corresponding to the far-field channel model and near-field channel model, respectively. The boundary of these two fields is determined by the Rayleigh distance~\cite{RayDistance}, which is proportional to the square of the array aperture. In the RIS assisted system, the array aperture is not very large, and the Rayleigh distance is small. The scatters are generally assumed in the far-field region of RIS. The existing codebook for beam training is designed based on the far-field channel model~\cite{RISBT,JunBT,DNBT}. With the increasing number of RIS elements from RIS to XL-RIS (e.g., from 256 to 1024), the array aperture of XL-RIS is very large, and the Rayleigh distance increases accordingly~\cite{Mingyao}. The scatters are more likely to be in the near-field region of XL-RIS, and the near-field channel model should be considered in the XL-RIS assisted system. The existing far-field codebook mismatches the near-field channel model. Thus, the corresponding far-field beam training will cause severe performance loss in the XL-RIS assisted near-field communications. Unfortunately, this important problem has not been studied in the literature.

To full in this gap, we propose the efficient near-field beam training schemes by designing the near-field codebook to match the near-field channel model in this paper\footnote{Simulation codes will be provided in the following link to reproduce the results presented in this paper after publication: http://oa.ee.tsinghua.edu.cn/dailinglong/publications/publications.html.}. Our contributions are summarized as follows.

\begin{enumerate}

\item We design the near-field codebook to match the near-field channel model, and then propose the corresponding near-field beam training scheme for XL-RIS. Specifically, by considering the near-field cascaded array steering vector of the XL-RIS cascaded channel, the near-field codebook is firstly designed, where each codeword is determined by a pair of sampled points in the $x\mbox{-}y\mbox{-}z$ coordinate system. Then, the optimal codeword for XL-RIS is obtained by the exhausted training procedure between the XL-RIS and the user.

\item In order to reduce the beam training overhead, we further design a hierarchical near-field codebook and propose the corresponding hierarchical near-field beam training scheme for XL-RIS. Compared with the near-field codebook, the hierarchical near-field codebook consists of several different levels of sub-codebooks, which are determined by different sampling ranges and sampling steps. During beam training, we search from the first level sub-codebook to the last level sub-codebook in turn, where the sampling ranges and sampling steps gradually become smaller. Finally, the globally optimal codeword can be obtained in the last level sub-codebook associated with the minimum sampling ranges and sampling steps.

\end{enumerate}

The rest of the paper is organized as follows. In Section II, we firstly introduce the signal model, and then review the existing far-field channel model and far-field codebook. The near-field channel model for the XL-RIS assisted system is also presented in Section II. In Section III, the near-field codebook is designed and the corresponding near-field beam training scheme is proposed, and then the hierarchical near-field codebook based beam training is further proposed to reduce the beam training overhead. Simulation results and conclusions are provided in Section IV and Section V, respectively.

{\it Notation}: Lower-case and upper-case boldface letters ${\bf{a}}$ and ${\bf{A}}$ denote a vector and a matrix, respectively; ${{{\bf{a}}^*}}$ and ${{{\bf{a}}^H}}$ denote the conjugate and conjugate transpose of vector $\bf{a}$, respectively; ${{\|{\bf{a}}\|_2}}$ denotes the $l_2$ norm of vector $\bf{a}$; ${{{\bf{A}}^{H}}}$ denotes the conjugate transpose of matrix $\bf{A}$. $\cal CN\left(\mu,\sigma \right)$ denotes the probability density function of the circularly symmetric complex Gaussian distribution with mean $\mu$ and variance $\sigma^2$. Finally, ${\rm{diag}}\left({\bf{a}}\right)$ denotes the diagonal matrix with the vector $\bf{a}$ on its diagonal.
\vspace{-1mm}

\section{System Model}\label{S2}
In this section, we will first introduce the signal model of the XL-RIS assisted communication system. Then, the existing far-field channel model and the far-field codebook for beam training will be briefly reviewed. Finally, the near-field channel model for XL-RIS is presented.

\subsection{Signal Model}\label{S2.1}

\begin{figure}[htbp]
\begin{center}
\includegraphics[width=0.5\linewidth]{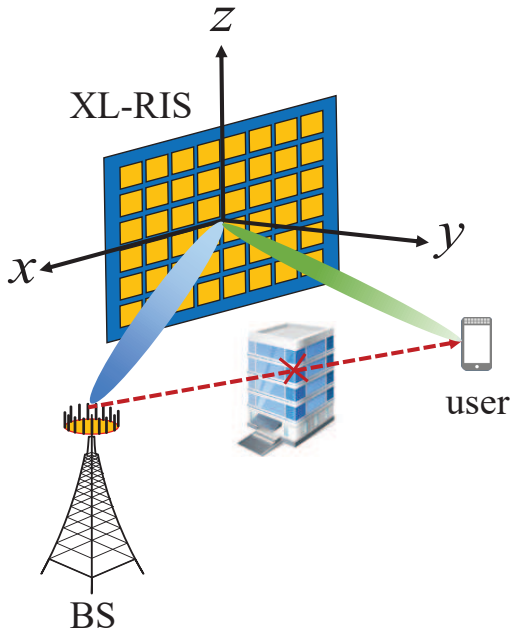}
\end{center}
\setlength{\abovecaptionskip}{-0.3cm}
\caption{The XL-RIS assisted wireless communication system.} \label{HF}
\end{figure}

As shown in Fig. 1, a XL-RIS is deployed between the BS with $M$-element antenna array and a single-antenna user to provide a reflecting link to assist communication, where the direct link between the BS and the user is blocked by obstacles~\cite{ChongweiCE,RISBT}. The XL-RIS consisting of $N=N_1\times N_2$ elements is placed in the $x\mbox{-}z$ plane, where the center of the XL-RIS is at the origin of the $x\mbox{-}y\mbox{-}z$ coordinate system.

Let ${\bf{G}}\in\mathbb{C}^{N\times M}$ denote the channel from the BS to the XL-RIS and ${\bf{h}}_{r}\in\mathbb{C}^{1\times N}$ denote the channel from the XL-RIS to the user. By considering the downlink transmission, the received signal $r$ at the user can be expressed by
\begin{equation}\label{eq1}
r = {\bf{h}}_{r}{\rm{diag}}\left({\bm{\theta}}\right){\bf{Gv}}s + n,
\end{equation}
where ${\bm{\theta}}=[\theta_{1},\cdots,\theta_{N}]\in\mathbb{C}^{N\times 1}$ is the reflecting beamforming vector at the XL-RIS with $\theta_{n}$ representing the reflecting coefficient at the $n$th RIS element $(n=1, \cdots,N)$, ${\bf{v}}\in\mathbb{C}^{M\times 1}$ represents the beamforming vector at the BS, $s$ represents the symbol transmitted by the BS, $n\sim{\cal C}{\cal N}\left( {0,\sigma^2}\right)$ represents the received noise at the user with ${\sigma^2}$ representing the noise power.

To design the effective beamforming vectors ${\bf{v}}$ and ${\bm{\theta}}$, it is necessary to acquire the accurate CSI~\cite{JunPrecoding,Zijian}. Since there are extremely large number of RIS elements, channel estimation cannot achieve the satisfactory channel estimation with low pilot overhead. By contrast, beam training is a more effective way to acquire CSI~\cite{RISBT,DNBT}. Specifically, since the XL-RIS is generally dominated by the main path (or a few paths), we only need to search the physical direction of the main path by beam training instead of explicitly estimating the entire channel. Thus, in the following description, only the main path is concerned, and the corresponding beam training method will be investigated to search the optimal directional beam to align with the main path.


Moreover, since the BS and the XL-RIS are generally deployed at fixed positions, the channel ${\bf{G}}$ from the BS to the XL-RIS has a much longer channel coherence time than the channel ${\bf{h}}_r$ from the XL-RIS to the user due to the user's mobility~\cite{RISBT}. For simplicity, we assume that the beamforming vector ${\bf{v}}$ at the BS has been aligned with the main path of the channel $\bf{G}$~\cite{RISBT}. In this paper, we only focus on the beam training at the XL-RIS. Next, we will briefly review the existing far-field channel model and the corresponding far-field codebook for beam training.

\subsection{Far-Field Channel Model and Far-Filed Codebook}\label{S2.2}

Based on the far-field channel model, $\bf{G}$ and ${\bf{h}}_r$ can be respectively represented by
\begin{equation}\label{eq2}
{\bf{G}}_{\rm{far\mbox{-}field}}={\alpha_{G}}{\bf{a}}\left({\phi}_{G_r},{\psi}_{G_r}\right){\bf{b}}^T\left({\phi}_{G_t},{\psi}_{G_t}\right),
\end{equation}
\begin{equation}\label{eq3}
{\bf{h}}^r_{\rm{far\mbox{-}field}}={\alpha_{r}}{\bf{a}}^T\left({\phi}_{r},{\psi}_{r}\right),
\end{equation}
where ${\alpha_{G}}$ and ${\alpha_{r}}$ represent the path gains, ${\phi}_{G_r}$ and ${\psi}_{G_r}$ represent the spatial angles at the XL-RIS for the channel $\bf{G}$, ${\phi}_{G_t}$ and ${\psi}_{G_t}$ represent the spatial angles at the BS for the channel $\bf{G}$, ${\phi}_{r}$ and ${\psi}_{r}$ represent the spatial angles at the XL-RIS for the channel ${\bf{h}}_r$. ${\bf{a}}(\phi, \psi)$ and ${\bf{b}}(\phi, \psi)$ represent the far-field array steering vector associated to the XL-RIS and the BS, respectively. Take ${\bf{a}}(\phi, \psi)$ as an example, it can be expressed as~\cite{PartII}
\begin{equation}\label{eq4}
{\bf{a}}\left(\phi,\psi \right) = {\left[ {{e^{ - j2{\pi}{\phi}{\bf{n}}_1}}}\right]}{\otimes}{\left[ {{e^{ - j2{\pi}{\psi} {\bf{n}}_2}}} \right]},
\end{equation}
where ${\bf{n}}_1=[0,1,\cdots,N_1-1]^T$ and ${\bf{n}}_2=[0,1,\cdots,N_2-1]^T$. $\phi = d_f{\rm{sin}}\left(\vartheta\right){\rm{cos}}\left(\upsilon\right)/{\lambda}$ and $\psi=d_f{\rm{sin}}\left(\upsilon\right)/{\lambda}$, where $\vartheta$ and $\upsilon$ respectively represent the physical angles in the azimuth and elevation, $\lambda$ is the carrier wavelength, and $d_f$ is the element spacing satisfying $d_f = \lambda/2$.

By considering that the beamforming vector $\bf{v}$ at the BS has been designed, i.e., ${\bf{v}}=\frac{{\bf{b}}^*}{\sqrt M}$, the receiver signal $r$ in~(\ref{eq1}) can be further represented by
\begin{equation}\label{eq5}
\begin{aligned}
 r= & {\alpha}{\bf{a}}^T\left({\phi}_{r},{\psi}_{r}\right){\rm{diag}}\left({\bm{\theta}}\right){\bf{a}}\left({\phi}_{G_r},{\psi}_{G_r}\right){\bar s} + n
\\ = & {\alpha}{\bm{\theta}}^T{\rm{diag}}\left({\bf{a}}\left({\phi}_{r},{\psi}_{r}\right)\right){\bf{a}}\left({\phi}_{G_r},{\psi}_{G_r}\right){\bar s} + n
\\ = & {\alpha}{\bm{\theta}}^T{\bf{a}}\left({\phi}_{G_r}+{\phi}_{r},{\psi}_{G_r}+{\psi}_{r}\right){\bar s} + n
\\ = & {\bm{\theta}}^T{\bar{{\bf{h}}}}_{\rm{far\mbox{-}field}}{\bar s} + n,
\end{aligned}
\end{equation}
where ${\bar{{\bf{h}}}}_{\rm{far\mbox{-}field}}={\alpha}{\bf{a}}\left({\phi}_{G_r}+{\phi}_{r},{\psi}_{G_r}+{\psi}_{r}\right)$ denotes the far-field cascaded channel, $\alpha={\alpha}_G{\alpha}_r$ denotes the effective gain of ${\bar{{\bf{h}}}}_{\rm{far\mbox{-}field}}$, and $\bar s={\bf{b}}^T\left({\phi}_{G_t},{\psi}_{G_t}\right){\bf{v}}s$ denotes the effective transmitted symbol.

For beam training, the entire procedure can be divided into multiple time slots. In different time slots, the reflecting beamforming vector ${\bm{\theta}}$ is set as different codewords in the predefined codebook, which will equivalently produce different directional beams. For each codeword, the user will measure the strength of the received signal $r$ and feedback the optimal codeword index. Based on the far-field array steering vector in~(\ref{eq4}), the existing far-field codebook ${\bf{F}}$ is generally designed as~\cite{RISBT}
\begin{equation}\label{eq6}
{\bf{F}}=\left[{\bf{a}}^*\left({\phi}_1,{\psi}_1\right),\cdots,{\bf{a}}^*\left({\phi}_1,{\psi}_{N_1}\right),\cdots,{\bf{a}}^*\left({\phi}_{N_1},{\psi}_1\right),\cdots,{\bf{a}}^*\left({\phi}_{N_1},{\psi}_{N_2}\right)\right],
\end{equation}
where ${\phi}_n={\frac{2n-N_1-1}{N_1}}$ with $n=1,2,\cdots,N_1$ and ${\psi}_n={\frac{2n-N_2-1}{N_2}}$ for ${n = 1,2, \cdots ,N_{2}}$. Each column of $\bf{F}$ represents a codeword for ${\bm{\theta}}$.

The existing beam training schemes are mainly based on the above far-field codebook ~\cite{RISBT,JunBT,DNBT}. However, when the RIS develops into XL-RIS, the far-field codebook and beam training may not be applicable any more, which will be explained in detail in the next Section II-C.


\subsection{Near-Field Channel Model}\label{S2.2}

\begin{figure}[htbp]
\begin{center}
\includegraphics[width=0.8\linewidth]{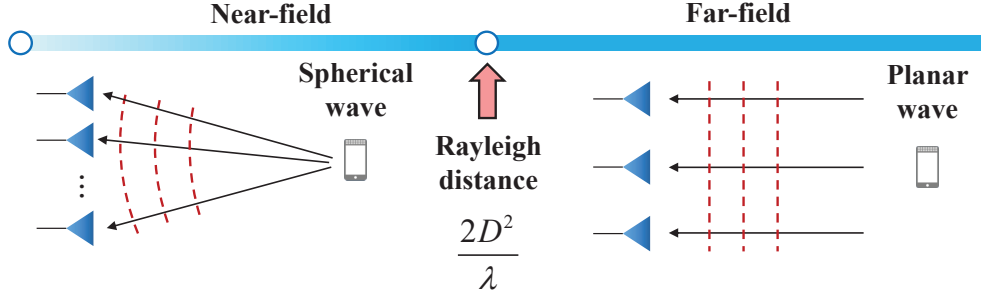}
\end{center}
\setlength{\abovecaptionskip}{-0.3cm}
\caption{The near-field region and the far-field region~\cite{Mingyao}.} \label{HF}
\end{figure}

Specifically, as shown in Fig. 2, the electromagnetic radiation field in wireless communications can be divided into far-field region and near-field region~\cite{RayDistance}, where different fields will result in different channel models. The bound between these two fields is determined by the Rayleigh distance $Z=\frac{2D^2}{\lambda}$, where $D$ represents the array aperture. In the conventional RIS assisted system, since the array aperture of the RIS is not too large, the corresponding Rayleigh distance is small. The scatters are in the far-field region of RIS~\cite{JunPrecoding,ChongweiCE,RISBT}, where the RIS channel can be modeled under the planar wave assumption, as described in~(\ref{eq2}) and~(\ref{eq3}). With the increase of the array aperture in the XL-RIS assisted system, the corresponding Rayleigh distance also increases. For example, we consider that the carrier frequency is $30$ GHz and the corresponding carrier wavelength is $\lambda=0.01$ meters. When the array aperture for RIS is $D=0.1$ meters, the Rayleigh distance is only $Z=2$ meters. When the array aperture for XL-RIS increases to $D=1$ meter, the Rayleigh distance can reach $Z=200$ meters. Thus, in the XL-RIS assisted system, the scatters are more likely to be in the near-field region, where the XL-RIS channel should be modeled under the spherical wave assumption. Next, we will introduce the near-field channel model of XL-RIS.

For convenience of description, the distances and coordinates are normalized by the carrier wavelength $\lambda$ in the following of this paper~\cite{JinShi}. The vertical or horizontal distance between two adjacent RIS elements is set to $d$. The coordinate of the RIS element can be represented as $((n_1-\frac{N_1+1}{2})d,0,(n_2-\frac{N_2+1}{2})d)$ in the $x\mbox{-}y\mbox{-}z$ coordinate system, where $n_1=1,\cdots,N_1$ and $n_2=1,\cdots,N_2$.

The near-field channel ${\bf{h}}^r_{\rm{near\mbox{-}field}}$ from the XL-RIS to the user can be represented by~\cite{JinShi}
\begin{equation}\label{eq7}
{\bf{h}}^r_{\rm{near\mbox{-}field}}={\alpha_{r}}{\bf{c}}^T\left(x_r,y_r,z_r\right),
\end{equation}
where $(x_r,y_r,z_r)$ represents the coordinate of the scatter corresponding to the main path between the XL-RIS and user. Compared with the far-field channel model in~(\ref{eq3}), the array steering vector ${\bf{c}}\left(x_r,y_r,z_r\right)$ for the near-field channel model is derived based on the the spherical wave assumption, which can represented by~\cite{JinShi}
\begin{equation}\label{eq8}
{\bf{c}}(x_r,y_r,z_r) = \left[e^{-j2\pi D^r(1,1)}, \cdots, e^{-j2\pi D^r(1,N_2)}, \cdots,e^{-j2\pi D^r(N_1,1)},\cdots,e^{-j2\pi D^r(N_1,N_2)}\right]^T,
\end{equation}
where $D^r(n_1,n_2) = \sqrt{(x_r-(n_1-\frac{N_1+1}{2})d)^2+y_r^2+(z_r-(n_2-\frac{N_2+1}{2})d)^2}$ represents the distance from the $(n_1,n_2)$-th RIS element to $(x_r,y_r,z_r)$.

Similarly, the array steering vector at the RIS of the channel $\bf{G}$ from the BS to the XL-RIS should be also near-field. Thus, the near-field cascaded channel ${\bar{{\bf{h}}}}_{\rm{near\mbox{-}field}}$ in~(\ref{eq5}) can be represented by
\begin{equation}\label{eq9}
{\bar{{\bf{h}}}}_{\rm{near\mbox{-}field}}={\alpha}{{\bar{\bf{c}}}}\left((x_{G_r},y_{G_r},z_{G_r}),(x_r,y_r,z_r)\right),
\end{equation}
where $(x_{G_r},y_{G_r},z_{G_r})$ represents the coordinate of the scatter corresponding to the main path between the BS and XL-RIS. The near-field cascaded array steering vector ${{\bar{\bf{c}}}}\left((x_{G_r},y_{G_r},z_{G_r}),(x_r,y_r,z_r)\right)$ can be represented by
\begin{equation}\label{eq10}
\begin{aligned}
{{\bar{\bf{c}}}}\left((x_{G_r},y_{G_r},z_{G_r}),(x_r,y_r,z_r)\right) = \left[e^{-j2\pi D(1,1)}, \cdots, e^{-j2\pi D(1,N_2)}, \cdots,\right.\\\left.e^{-j2\pi D(N_1,1)},\cdots,e^{-j2\pi D(N_1,N_2)}\right]^T,
\end{aligned}
\end{equation}
where $D(n_1,n_2)=D^r(n_1,n_2) + D^{G_r}(n_1,n_2)$ represents the effective distance of ${\bar{{\bf{h}}}}_{\rm{near\mbox{-}field}}$, and $D^{G_r}(n_1,n_2)=\sqrt{(x_{G_r}-(n_1-\frac{N_1+1}{2})d)^2+y_{G_r}^2+(z_{G_r}-(n_2-\frac{N_2+1}{2})d)^2}$. $\left(x_{G_r},y_{G_r},z_{G_r}\right)$ satisfies $X_{\rm{min}}^{G_r}\leq x_{G_r}\leq X_{\rm{max}}^{G_r}$, $Y_{\rm{min}}^{G_r}\leq y_{G_r}\leq Y_{\rm{max}}^{G_r}$, $Z_{\rm{min}}^{G_r}\leq z_{G_r}\leq Z_{\rm{max}}^{G_r}$, and $\left(x_{r},y_{r},z_{r}\right)$ satisfies $X_{\rm{min}}^{r}\leq x_{r}\leq X_{\rm{max}}^{r}$,$Y_{\rm{min}}^{r}\leq y_{r}\leq Y_{\rm{max}}^{r}$, and $Z_{\rm{min}}^{r} \leq z_{r}\leq Z_{\rm{max}}^{r}$.

Compared with the far-field cascaded array steering vector only associated with the angles in~(\ref{eq4}), the near-field cascaded array steering vector is determined by a pair of points in the $x\mbox{-}y\mbox{-}z$ coordinate system, i.e., $(x_{G_r},y_{G_r},z_{G_r})$ and $(x_r,y_r,z_r)$. The existing far-field codebook mismatches the near-field channel model. Thus, the corresponding far-field beam training will cause severe performance loss in the XL-RIS assisted near-field communications. In this paper, we design the near-field codebook to match the near-field channel model, and then propose the corresponding near-field beam training for XL-RIS, which will be introduced in the next Section III.

\section{Proposed Near-Field Codebook Design and Beam Training Scheme for XL-RIS}\label{S3}
In this section, we will introduce the proposed near-field codebook and the corresponding near-field beam training. Then, a hierachical near-field codebook and the corresponding beam training will be further proposed to reduce the beam training overhead.

\subsection{Proposed Near-Field Codebook Design and Beam training}\label{S3.1}

Inspired by the near-field dictionary matrix design for the XL-MIMO channel estimation~\cite{JinShi}, we design a near-field codebook for the XL-RIS beam training. In the near-field dictionary matrix proposed in~\cite{JinShi} for the extremely large-scale linear antenna array, the considered entire two-dimensional (2D) plane is divided by several sampled points in the $x\mbox{-}y$ coordinate system. Each column of the dictionary matrix is the corresponding near-field array steering vector for the linear array associated with one sampled point. If the near-field codebook for the planar array is required to design in the XL-MIMO system, we only need to extend the 2D plane to the three-dimensional (3D) space, where each codeword can be generated based on the near-field array steering vector for the planar array associated with one sampled point in the $x\mbox{-}y\mbox{-}z$ coordinate system. However, the near-field cascaded channel brings new challenges to the codebook design for XL-RIS .

Specifically, since the near-field cascaded array steering vector ${{\bar{\bf{c}}}}$ of the near-field cascaded channel ${\bar{{\bf{h}}}}_{\rm{near\mbox{-}field}}$ is determined by the sum of the distance from $\left(x_{G_r},y_{G_r},z_{G_r}\right)$ to the XL-RIS and the distance from $\left( x_r,y_r,z_r\right)$ to the XL-RIS, each codeword for XL-RIS should be related to a pair of sampled points in the $x\mbox{-}y\mbox{-}z$ coordinate system, instead of only one sampled point~\cite{JinShi}. Next, we will introduce the designed near-field codebook based on the near-field cascaded array steering vector.

\subsubsection{Near-Field Codebook Design}
Let $\Xi^{G_r}$ and $\Xi^{r}$ denote the two collections of sampled points corresponding to $\left(x_{G_r},y_{G_r},z_{G_r}\right)$ and $\left( x_r,y_r,z_r\right)$, which can be represented as
\begin{equation}\label{eq11}
\begin{aligned}
{\Xi}^{G_r} = \left\{(x^{G_r}_s,y^{G_r}_s,z^{G_r}_s)|x^{G_r}_s=X^{G_r}_{\rm{min}},X^{G_r}_{\rm{min}}+\Delta x^{G_r},\cdots,X^{G_r}_{\rm{max}};y^{G_r}_s=Y^{G_r}_{\rm{min}},\right.\\\left.Y^{G_r}_{\rm{min}}+\Delta y^{G_r},\cdots,Y^{G_r}_{\rm{max}}; z^{G_r}_s=Z^{G_r}_{\rm{min}},Z^{G_r}_{\rm{min}}+\Delta z^{G_r},\cdots,Z^{G_r}_{\rm{max}}\right\},
\end{aligned}
\end{equation}
\begin{equation}\label{eq12}
\begin{aligned}
{\Xi}^{r} = \left\{(x^{r}_s,y^{r}_s,z^{r}_s)|x^{r}_s=X^{r}_{\rm{min}},X^{r}_{\rm{min}}+\Delta x^{r},\cdots,X^{r}_{\rm{max}};y^{r}_s=Y^{r}_{\rm{min}},\right.\\\left.Y^{r}_{\rm{min}}+\Delta y^{r},\cdots,Y^{r}_{\rm{max}}; z^{r}_s=Z^{r}_{\rm{min}},Z^{r}_{\rm{min}}+\Delta z^{r},\cdots,Z^{r}_{\rm{max}}\right\},
\end{aligned}
\end{equation}
where $\Delta x^{G_r}$, $\Delta y^{G_r}$ and $\Delta z^{G_r}$ represent the sampling step on the $x$-axis, $y$-axis and $z$-axis for ${\Xi}^{G_r}$, respectively. $\Delta x^{r}$, $\Delta y^{r}$ and $\Delta z^{r}$ represent the sampling step on the $x$-axis, $y$-axis and $z$-axis for ${\Xi}^{r}$, respectively. Let $\Delta =[\Delta x^{G_r},\Delta y^{G_r},\Delta z^{G_r},\Delta x^{r},\Delta y^{r},\Delta z^{r}]$ denote all sampling steps. Given a pair of sampled points $(x^{G_r}_s,y^{G_r}_s,z^{G_r}_s)$ and $(x^{r}_s,y^{r}_s,z^{r}_s)$, the effective sampled distance $D_s(n_1,n_2)$ can be presented by
\begin{equation}\label{eq13}
\begin{aligned}
D_s(n_1,n_2) = & \sqrt{\bigg(x^{G_r}_s-(n_1-\frac{N_1+1}{2})d\bigg)^2+{y^{G_r}_s}^2+\bigg(z^{G_r}_s-(n_2-\frac{N_2+1}{2})d\bigg)^2}\\
+ & \sqrt{\bigg(x^{r}_s-(n_1-\frac{N_1+1}{2})d\bigg)^2+{y^{r}_s}^2+\bigg(z^{r}_s-(n_2-\frac{N_2+1}{2})d\bigg)^2},
\end{aligned}
\end{equation}
where $n_1=1,2,\cdots,N_1$ and $n_2=1,2,\cdots,N_2$.

\begin{algorithm}[htbp]
\caption{Near-field codebook design}
\textbf{Inputs}: The two collections of sampled points $\Xi^{G_r}$ and $\Xi^{r}$, the number of RIS elements $N_1$ and $N_2$.
\\\textbf{Initialization}: ${\bf{W}}=\emptyset$, $L=0$.
\\1. \textbf{for} $(x^{G_r}_s,y^{G_r}_s,z^{G_r}_s)\in \Xi^{G_r}$ \textbf{do}
\\2. \hspace*{+3mm}\textbf{for} $(x^{r}_s,y^{r}_s,z^{r}_s)\in \Xi^{r}$ \textbf{do}
\\3. \hspace*{+6mm}${\bar{\bf{c}}}_s=\left[e^{-j2\pi D_s(1,1)}, \cdots, e^{-j2\pi D_s(1,N_2)}, \cdots,e^{-j2\pi D_s(N_1,1)},\cdots,e^{-j2\pi D_s(N_1,N_2)}\right]^H$
\\4. \hspace*{+6mm}\textbf{if} ${\bf{w}}\notin {\bf{W}}$ \textbf{then}
\\5. \hspace*{+9mm}${\bf{W}}=[{\bf{W}},{\bar{\bf{c}}}_s]$
\\6. \hspace*{+9mm}$L=L+1$
\\7. \hspace*{+6mm}\textbf{end if}
\\8. \hspace*{+3mm}\textbf{end for}
\\9. \textbf{end for}
\\\textbf{Output}: The designed near-field XL-RIS codebook ${\bf{W}}$, and the codebook size $L$.
\end{algorithm}

\textbf{Algorithm 1} shows the specific near-field codebook design procedure. From Step 3, we can find that the near-field codeword is generated based on the near-field cascaded array steering vector, which is related to a pair of sampled points $(x^{G_r}_s,y^{G_r}_s,z^{G_r}_s)$ and $(x^{r}_s,y^{r}_s,z^{r}_s)$. It is noted that different pairs of sampled points may produce the same effective sampled distance, which will result in the same codeword. In order to solve this problem, we need to ensure that each new codeword is different from all previous codewords, as shown in Steps 4-7. Finally, the designed near-field codebook ${\bf{W}}$ is obtained, where each column represents one codeword for the reflecting beamforming vector $\bm{\theta}$ at the XL-RIS. The corresponding codebook size $L$ is also obtained.

After designing the near-field codebook for XL-RIS, the beam training procedure between the XL-RIS and the user can be performed to search the optimal codeword for the reflecting beamforming vector $\bm{\theta}$ at the XL-RIS. Next, we will introduce the corresponding near-field beam training scheme.

\subsubsection{Near-Field Beam Training}

The specific near-field beam training procedure is summarized in \textbf{Algorithm 2}, where all the codewords in the designed near-field codebook ${\bf{W}}$ need to be traversed. The entire training procedure can be divided into $L$ time slots. In $l$-th time slots, the BS transmits the effective symbol $\bar{s}$ to the user, where the reflecting beamforming vector ${\bm{\theta}}_l$ is set as the $l$-th codeword in the designed near-field codebook ${\bf{W}}$ at the XL-RIS, as shown in Step 3. After $L$ time slots, the user can obtain the optimal codeword based on all received signals $\{r_l\}_{l=1}^{L}$ with the help of Steps 4-7. Finally, the optimal codeword index $l_{\rm{opt}}$ is fed back from the user to XL-RIS.

\begin{algorithm}[htbp]
\caption{Near-field beam training}
\textbf{Inputs}: The designed near-field XL-RIS codebook ${\bf{W}}$, and the effective transmitted
symbol $\bar{s}$.
\\\textbf{Initialization}: $l=0$, $|r|_{\rm{opt}}=0$, $l_{\rm{opt}}=0$.
\\1. \textbf{for} ${{\bar{\bf{c}}}_s}\in {\bf{W}}^H$ \textbf{do}
\\2. \hspace*{+3mm} $l=l+1$
\\3. \hspace*{+3mm} ${r}_l= {\bm{\theta}}_l^T{\bar{{\bf{h}}}}_{\rm{near\mbox{-}field}}{\bar s} + n_l$, where ${\bm{\theta}}_l={{\bar{\bf{c}}}_s}$
\\4. \hspace*{+3mm} \textbf{if} $|r_l|>|r|_{\rm{opt}}$ \textbf{then}
\\5. \hspace*{+6mm} $l_{\rm{opt}}=l$
\\6. \hspace*{+6mm} $|r|_{\rm{opt}}=|r_l|$
\\7. \hspace*{+3mm} \textbf{end if}
\\8. \textbf{end for}
\\\textbf{Output}: The feedback optimal codeword index $l_{\rm{opt}}$ from the user.
\end{algorithm}

Since the near-field cascaded array steering vector for XL-RIS is jointly determined by a pair of sampled points, the codebook size $L$ is usually large. This exhausted training procedure will lead to huge beaming training overhead. In order to reduce the beam training overhead, we further design a hierachical near-field codebook and propose the corresponding hierachical beam training scheme.

\subsection{Proposed Hierachical Near-Field Codebook Design and Beam training}\label{S3.1}

To reduce the beam training overhead, one effective way is to reduce the codebook size. By referring to~(\ref{eq11}) and~(\ref{eq12}), we can find the codebook size for the entire sampling space is mainly determined by the sampling steps on the $x$-axis, $y$-axis and $z$-axis. If the sampling steps are increased, the codebook size will be reduced. But the performance of the beam training will be also degraded accordingly, since it is difficult to accurately locate the scatters corresponding to the main paths with the reduced codebook. In order to solve this problem, we design a hierachical near-field codebook, which consists of several different levels of sub-codebooks. These different levels of sub-codebooks are determined by different sampling ranges and sampling steps.

Specifically, let $K$ denote the number of different levels of sub-codebooks. In the $k$-th sub-codebook ($k=1,2,\cdots,K$), the corresponding collections of sampled points ${\Xi}^{G_r}_k$ and ${\Xi}^{r}_k$ can be defined as
\begin{equation}\label{eq13}
\begin{aligned}
{\Xi}^{G_r}_k = \left\{(x^{G_r,k}_s,y^{G_r,k}_s,z^{G_r,k}_s)|x^{G_r,k}_s=X^{G_r,k}_{\rm{min}},X^{G_r,k}_{\rm{min}}+\Delta x^{G_r,k},\cdots,X^{G_r,k}_{\rm{max}};y^{G_r,k}_s=Y^{G_r,k}_{\rm{min}},\right.\\\left.Y^{G_r,k}_{\rm{min}}+\Delta y^{G_r,k},\cdots,Y^{G_r,k}_{\rm{max}}; z^{G_r,k}_s=Z^{G_r,k}_{\rm{min}},Z^{G_r,k}_{\rm{min}}+\Delta z^{G_r,k},\cdots,Z^{G_r,k}_{\rm{max}}\right\},
\end{aligned}
\end{equation}
\begin{equation}\label{eq14}
\begin{aligned}
{\Xi}^{r}_k = \left\{(x^{r,k}_s,y^{r,k}_s,z^{r,k}_s)|x^{r,k}_s=X^{r,k}_{\rm{min}},X^{r,k}_{\rm{min}}+\Delta x^{r,k},\cdots,X^{r,k}_{\rm{max}};y^{r,k}_s=Y^{r,k}_{\rm{min}},\right.\\\left.Y^{r,k}_{\rm{min}}+\Delta y^{r,k},\cdots,Y^{r,k}_{\rm{max}}; z^{r,k}_s=Z^{r,k}_{\rm{min}},Z^{r,k}_{\rm{min}}+\Delta z^{r,k},\cdots,Z^{r,k}_{\rm{max}}\right\}.
\end{aligned}
\end{equation}
Take the sampling on the $x$-axis for ${\Xi}^{G_r}_k$ as an example, $[X^{G_r,k}_{\rm{min}},X^{G_r,k}_{\rm{max}}]$ and $\Delta x^{G_r,k}$ represent the sampling range and sampling step, respectively.
Further, let $R^{k}=\big\{[X^{G_r,k}_{\rm{min}},X^{G_r,k}_{\rm{max}}],[Y^{G_r,k}_{\rm{min}},Y^{G_r,k}_{\rm{max}}],\\{[Z^{G_r,k}_{\rm{min}},Z^{G_r,k}_{\rm{max}}]},[X^{r,k}_{\rm{min}},X^{r,k}_{\rm{max}}],[Y^{r,k}_{\rm{min}},Y^{r,k}_{\rm{max}}],{[Z^{r,k}_{\rm{min}},Z^{r,k}_{\rm{max}}]}\big\}$
and ${\Delta}^{k}=\{\Delta x^{G_r,k},\Delta y^{G_r,k},\Delta z^{G_r,k},\\\Delta x^{r,k},\Delta y^{r,k},\Delta z^{r,k}\}$ respectively denote all sampling ranges and all sampling steps for the $k$-th level sub-codebook. In this way, ${\Xi}^{G_r}_k$ and ${\Xi}^{r}_k$ can be completely determined by $R^{k}$ and ${\Delta}^{k}$. Given ${\Xi}^{G_r}_k$ and ${\Xi}^{r}_k$, the $k$-th level sub-codebook ${\bf{W}}_k$ and the corresponding codebook size $L_k$ can be further obtained by referring to \textbf{Algorithm 1}. Fig. 3 shows the comparison between the near-field codebook and the hierachical near-field codebook. In the hierachical near-field codebook, from the $1$-st level sub-codebook to the $K$-th level sub-codebook, both the corresponding sampling ranges and sampling steps gradually become smaller. Thus, the codebook size of each level of sub-codeboook is not large.

\begin{figure}[htbp]
\begin{center}
\includegraphics[width=0.8\linewidth]{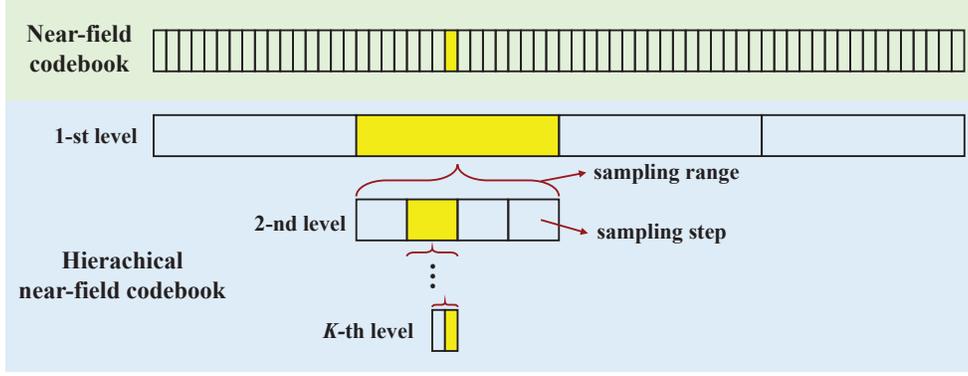}
\end{center}
\setlength{\abovecaptionskip}{-0.3cm}
\caption{Comparison between the near-field codebook and the hierachical near-field codebook.}
\end{figure}

Based on the hierachical near-field codebook, we further propose the hierachical near-field beam training scheme. The basic idea is to search from the $1$-st level sub-codebook to the $K$-th level sub-codebook in turn, where the sampling ranges of the latter level sub-codebook is determined by the optimal codeword searched and the sampling steps by the former level sub-codebook, as shown in Fig. 3. By assuming that the searched optimal codeword for the $k$-th level sub-codebook is ${{\bar{\bf{c}}}_{s,k,\rm{opt}}}$ with the  corresponding sampled points $(x^{G_r}_{s,k,\rm{opt}},y^{G_r}_{s,k,\rm{opt}},z^{G_r}_{s,k,\rm{opt}})$ and $(x^{r}_{s,k,\rm{opt}},y^{r}_{s,k,\rm{opt}},z^{r}_{s,k,\rm{opt}})$, the sampling ranges of the $(k+1)$-th level sub-codebook can be determined accordingly. Take $[X^{G_r,k+1}_{\rm{min}},X^{G_r,k+1}_{\rm{max}}]$ as an example, it can be represented by
\begin{equation}\label{eq15}
[X^{G_r,k+1}_{\rm{min}},X^{G_r,k+1}_{\rm{max}}] = [x^{G_r}_{s,k,\rm{opt}}-\Delta x^{G_r,k}/2, x^{G_r}_{s,k,\rm{opt}}+\Delta x^{G_r,k}/2].
\end{equation}

For the $1$-st level sub-codebook, the corresponding sampling ranges can be set as $R^{1}=\big\{[X^{G_r}_{\rm{min}},X^{G_r}_{\rm{max}}],[Y^{G_r}_{\rm{min}},Y^{G_r}_{\rm{max}}],{[Z^{G_r}_{\rm{min}},Z^{G_r}_{\rm{max}}]},[X^{r}_{\rm{min}},X^{r}_{\rm{max}}],[Y^{r}_{\rm{min}},Y^{r}_{\rm{max}}],[Z^{r,k}_{\rm{min}},Z^{r,k}_{\rm{max}}]\big\}$.
The initial sampling steps $\Delta^1$ should be set as bigger values to reduce the codebook size. In this paper, we set $\Delta^1=A\Delta$, where $\Delta$ is the sampling steps for the designed near-field codebook in Section III-A, and $A$ is a scalar greater than $1$. Moreover, we define a step control parameter $\delta$ ($0<\delta<1$) to gradually decrease the sampling steps of the sub-codebooks.

\begin{algorithm}[htbp]
\caption{Hierachical near-field beam training}
\textbf{Inputs}: The number of different levels of sub-codebooks $K$, the initial sampling ranges $R^1$, the initial sampling steps $\Delta^1$, the step control parameter $\delta$, and the effective transmitted symbol $\bar{s}$, the number of RIS elements $N_1$ and $N_2$.
\\\textbf{Initialization}: $l_k=0$, $|r|_{k,\rm{opt}}=0$ and $l_{k,\rm{opt}}=0$ for $\forall k$
\\1. \textbf{for} $k=1,2,\cdots, K$ \textbf{do}
\\2. \hspace*{+3mm} generate ${\Xi}^{G_r}_k$ and ${\Xi}^{r}_k$ based on $R^k$ and $\Delta^k$ by~(\ref{eq13}) and~(\ref{eq14})
\\3. \hspace*{+3mm} generate ${\bf{W}}_k$ based on ${\Xi}^{G_r}_k$ and ${\Xi}^{r}_k$ by \textbf{Algorithm 1}
\\4. \hspace*{+3mm} \textbf{for} ${{\bar{\bf{c}}}_{s,k}}\in {\bf{W}}_k$ \textbf{do}
\\5. \hspace*{+6mm} $l_k=l_k+1$
\\6. \hspace*{+6mm} ${r}_{l,k}= {\bm{\theta}}_{l,k}^T{\bar{{\bf{h}}}}_{\rm{near\mbox{-}field}}{\bar s} + n_{l,k}$, where ${\bm{\theta}}_{l,k}={{\bar{\bf{c}}}_{s,k}}$
\\7. \hspace*{+6mm} \textbf{if} $|r_{l,k}|>|r|_{k,\rm{opt}}$ \textbf{then}
\\8. \hspace*{+9mm} $l_{k,{\rm{opt}}}=l_k$
\\9.\hspace*{+9mm} $|r|_{k,\rm{opt}}=|r_{l,k}|$
\\10.\hspace*{+6mm} \textbf{end if}
\\11.\hspace*{+3mm} \textbf{end for}
\\12.\hspace*{+3mm} generate $R^{k+1}$ based on $l_{k,\rm{opt}}$ by~(\ref{eq15})
\\13.\hspace*{+3mm} $\Delta^{k+1}=\delta\Delta^{k}$
\\14.\textbf{end for}
\\\textbf{Output}: The feedback optimal codeword index $l_{K,{\rm{opt}}}$ from the user.
\end{algorithm}

The specific hierachical near-field beam training procedure is summarized in \textbf{Algorithm 3}, where the entire beam training procedure is divided into $K$ stages. In the $k$-th stage, the $k$-th level sub-codebook ${\bf{W}}_k$ is firstly generated by Steps 2-3. Then, Steps 4-11 are performed to search the optimal codeword in ${\bf{W}}_k$. It is noted that the searched optimal codeword index $l_{k,\rm{opt}}$ should be fed back from the user to the XL-RIS to generate the the sampling ranges $R^{k+1}$ and sampling steps $\Delta^{k+1}$ for the $(k+1)$-th level sub-codebook, as shown in Steps 12-13. Finally, the optimal codeword for the $K$-th level sub-codebook is regarded as the searched globally optimal codeword in the hierachical near-field codebook.

From \textbf{Algorithm 3}, we can find that the corresponding beam training overhead is $\sum_{k=1}^{K}L_k$. Since the sampling ranges and sampling steps for the hierachical near-field codebook are carefully designed, $L_k$ for $\forall k$ can be much less than $L$, which will be verified by the following simulation results.

%

\section{Simulation Results}\label{S5}

In this section, we provide the simulation results to verify the performance of the two proposed near-field beam training schemes.

For simulations, we consider that the number of antennas at the BS is $M=64$, and the number of RIS elements at the XL-RIS is $N=512$ ($N_1=128$ and $N_2=4$). The path gains are generated by ${\alpha _{G} \sim {\cal C}{\cal N}\left( {0,1} \right)}$ and ${\alpha _{r} \sim {\cal C}{\cal N}\left( {0,1} \right)}$. The vertical or horizontal distance between two adjacent RIS elements is set as $d=1/2$~\cite{JinShi}. The distance from the XL-RIS to the scatter is limited to $X^{G_r}_{\rm{min}}=X^{r}_{\rm{min}}=-1200d$, $X^{G_r}_{\rm{max}}=X^{r}_{\rm{max}}=1200d$, $Y^{G_r}_{\rm{min}}=Y^{r}_{\rm{min}}=10d$, $Y^{G_r}_{\rm{max}}=Y^{r}_{\rm{max}}=200d$, $Z^{G_r}_{\rm{min}}=Z^{r}_{\rm{min}}=-400d$, $Z^{G_r}_{\rm{max}}=Z^{r}_{\rm{max}}=-400d$. The symbol transmitted by the BS is set as $s=1$, and the beamforming vector at the BS is set as ${\bf{v}}=\frac{{\bf{b}}^*}{\sqrt M}$. Thus, the effective transmitted symbol is $\bar s=1$. The SNR is defined as $1/{\sigma}^2$.

We compare the proposed near-field beam training scheme and the hierachical near-field beam training scheme with the existing far-field beam training scheme~\cite{RISBT}. In the near-field beam training scheme, the sampling steps on all $x$-axis, $y$-axis and $z$-axis are set to the same, i.e., $\Delta x^{G_r}=\Delta y^{G_r}=\Delta z^{G_r}=\Delta x^{r}=\Delta y^{r}=\Delta z^{r}=\Delta_s$. That is to say $\Delta=[\Delta_s,\Delta_s,\Delta_s,\Delta_s,\Delta_s,\Delta_s]$. In the hierachical near-field beam training scheme, we set $A=4$, and the initial sampling steps $\Delta^1=A\Delta$. The step control parameter is set as $\delta=0.25$. The number of different levels of sub-codebooks is set as $K=2$. In the far-field beam training scheme, the far-field codebook $\bf{F}$ defined in~(\ref{eq3}) is adopted. Moreover, we provide the beamforming scheme with perfect CSI as the upper bound of performance, i.e., ${\bm{\theta}}=\frac{{\bar{\bf{c}}}^*}{\sqrt N}$.

\begin{figure}[htpb]
\begin{center}
\includegraphics[width=0.8\linewidth]{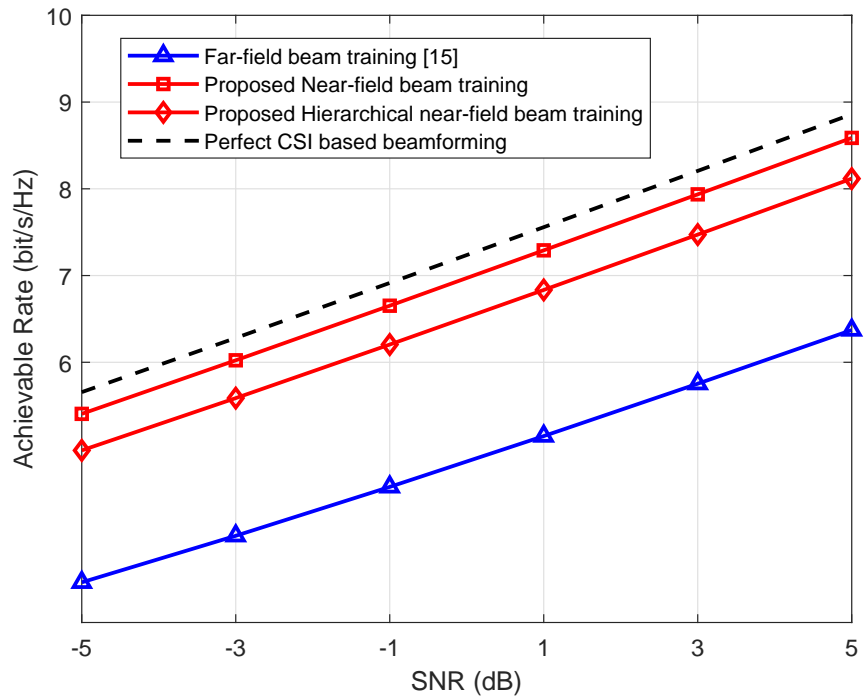}
\end{center}
\setlength{\abovecaptionskip}{-0.2cm}
\caption{Achievable rate performance comparison against the SNR.} \label{FIG3}
\vspace{-1mm}
\end{figure}

\begin{figure}[htpb]
\begin{center}
\includegraphics[width=0.8\linewidth]{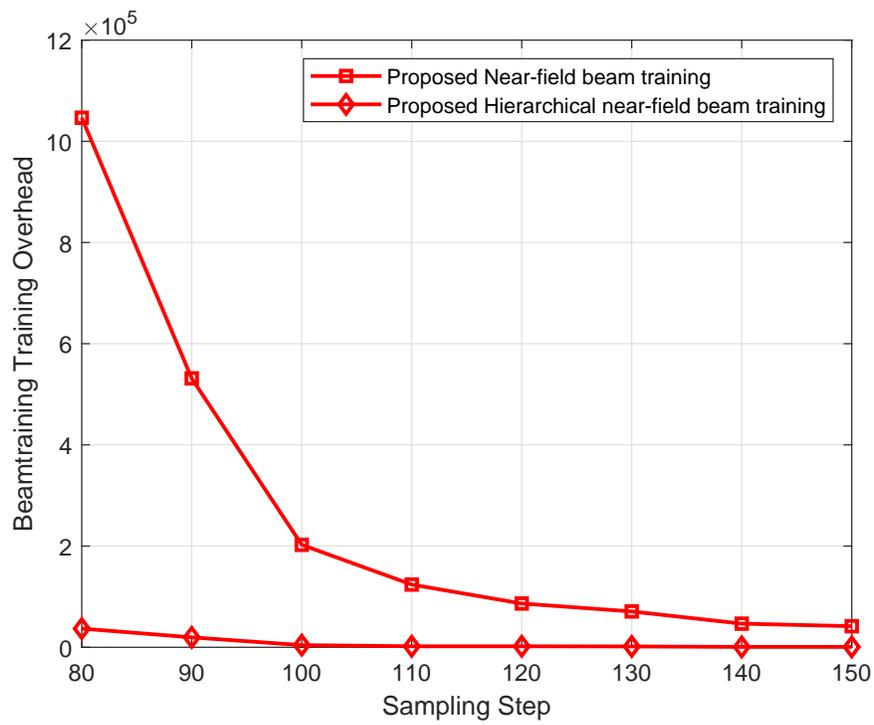}
\end{center}
\setlength{\abovecaptionskip}{-0.2cm}
\caption{Beam training overhead comparison against the sampling step $\Delta_s$.} \label{FIG3}
\vspace{-1mm}
\end{figure}

Fig. 4 shows the achievable rate performance comparison against the SNR, where $\Delta_s=100d$. We can find that compared with the existing far-field beam training scheme, the two proposed near-field beam training schemes can achieve better achievable rate performance. Due to the error propagation among different levels of sub-codebooks search, the performance of the hierachical near-field beam training scheme is sightly worse than that of the near-field beam training scheme. Specifically, the hierachical near-field beam training scheme can achieve about $92\%$ achievable rate performance of the near-field beam training scheme.

Fig. 5 further shows the beam training overhead comparison of the two proposed near-field beam training schemes against the sampling step $\Delta_s$. We can find that the hierachical near-field beam training scheme can greatly reduce the beam training overhead. When $\Delta_s=100d$, the beam training overhead of the near-field beam training scheme and the hierachical near-field beam training scheme are respectively $147628$ and $15927$, where the latter is only about $10\%$ of the former.

\section{Conclusions}\label{S6}
In this paper, we have proposed the two near-field beam training schemes by designing the near-field codebook for the XL-RIS assisted system. Simulation results shows that the two proposed near-field beam training schemes can achieve better performance than the existing far-field beam training scheme. Particulary, compared with the near-field beam training scheme, the hierachical near-field beam training scheme can reduce the beam training overhead by about $90\%$ with about $92\%$ achievable rate performance. For future works, the multi-beam training method can be used to the near-field XL-RIS beam training to further reduce the beam training overhead.

\bibliography{IEEEabrv,Near_Field_BT}

\end{document}